\documentclass[preprint, superscriptaddress,amsmath, nofootinbib]{revtex4-1}
\usepackage{graphicx}
\usepackage{latexsym}
\usepackage{slashed}
\usepackage{bm}
\usepackage{color}
\usepackage[colorlinks,citecolor=blue,urlcolor=blue,linkcolor=red]{hyperref}
\usepackage{diagbox}
\usepackage{makecell}
\pdfoutput=1
\usepackage{dcolumn}
\usepackage{bm}
\usepackage{multirow}
\usepackage{subcaption}
\usepackage{ulem}

\begin{document}

\title{Exploring interference effects between two ALP effective operators at the LHC}
\def\slash#1{#1\!\!\!/}

\author{Kingman Cheung}
\email{cheung@phys.nthu.edu.tw}
\affiliation{Department of Physics, National Tsing Hua University, Hsinchu 30013, Taiwan} 
\affiliation{Center for Theory and Computation,
National Tsing Hua University, Hsinchu 30013, Taiwan}
\affiliation{Division of Quantum Phases and Devices, School of Physics, Konkuk University, Seoul 143-701, Republic of Korea}

\author{Chih-Ting Lu}
\email{ctlu@njnu.edu.cn}
\affiliation{Department of Physics and Institute of Theoretical Physics, Nanjing Normal University, Nanjing, 210023, P. R. China} 

\author{C.J. Ouseph}
\email{ouseph444@gmail.com}
\affiliation{Department of Physics, National Tsing Hua University, Hsinchu 30013, Taiwan} 
\affiliation{Center for Theory and Computation,
National Tsing Hua University, Hsinchu 30013, Taiwan} 

\author{Priyanka Sarmah}
\email{sarmahpriyanka@gapp.nthu.edu.tw}
\affiliation{Department of Physics, National Tsing Hua University, Hsinchu 30013, Taiwan} 
\affiliation{Center for Theory and Computation,
National Tsing Hua University, Hsinchu 30013, Taiwan}

\date{\today}

\begin{abstract}
We observe that most studies of axion-like particle (ALP) production channels at the Large Hadron Collider (LHC) focus on a single type of ALP operator for each process in the effective field theory framework. In this work, we propose an alternative approach that considers two or more types of relevant ALP effective operators together in some specific ALP production channels and study their interference effects. Using the $p p\rightarrow t j a$ process with $a\rightarrow\gamma\gamma$ as an example, 
we show that this approach allows us to constrain the ALP interactions with both the $W$ boson and 
the top quark, as well as their interference in a single process. 
For the final state with two isolated photons and a top quark decaying semi-leptonically, 
we predict that the future bounds on the ALP decay constant can reach around 
$f_a \sim 10\;(20) $ TeV for $25$ GeV $< M_a < 100$ GeV at the LHC with 300 (3000) fb$^{-1}$ luminosity.
\end{abstract}

\maketitle
\section{Introduction}\label{sec.1}

Axion-like particles (ALPs) are hypothetical particles that arise in several extensions of the Standard Model (SM) of particle physics as  pseudo-Goldstone bosons with a shift symmetry~\cite{Preskill:1982cy,Abbott:1982af,Dine:1982ah,Bagger:1994hh,Svrcek:2006yi,Arvanitaki:2009fg,Cicoli:2012sz,Visinelli:2018utg,Chang:1999si,Bastero-Gil:2021oky}. They are characterized by a low mass and a weak coupling to ordinary matter, which makes them difficult to detect. However, their existence is predicted by various theories that attempt to solve some of the outstanding problems in physics, such as the 
strong CP problem~\cite{Peccei:1977hh,Peccei:1977ur,Weinberg:1977ma,Wilczek:1977pj,Kim:1979if,Shifman:1979if,Dine:1981rt}, the nature of dark matter~\cite{Duffy:2009ig,Chadha-Day:2021szb} and experimental anomalies~\cite{Chang:2021myh,Yuan:2022cpw,Liu:2022tqn,Cheung:2024kml}. 

In the effective field theory (EFT) framework~\cite{ALPmodel:2017}, possible interaction operators of ALP(s) 
and SM particles start at dimension-five~\cite{Georgi:1986df} and continue to 
higher order ones~\cite{Bauer:2018uxu}. 
We can explore the effect of each term separately or some of them collectively. On the other hand, 
the ALP mass range in the general EFT extends from almost massless to the electroweak scale or above. 
Therefore, there are various search strategies for different ALP interactions and masses including 
laboratory-based~\cite{Sikivie:1983ip}, beam-dump~\cite{Riordan:1987aw,Dobrich:2019dxc}, Higgs factories~\cite{Bauer:2017ris, Cheung:2023nzg},
and high energy collider~\cite{Bauer:2017ris,Bauer:2018uxu} experiments, as well as cosmological and astrophysical 
observations~\cite{CAST:2017uph} (for a recent summary please see Ref.~\cite{OHare:2024nmr}).

Among various kinds of ALP searches, the collider experiments are sensitive to probe the 
GeV to TeV scale ALPs. Although there were already a number of studies to explore the properties of 
ALPs at the Large Hadron 
Collider(LHC)~\cite{Jaeckel:2015jla,Brivio:2017ije,Bauer:2017ris,Ebadi:2019gij,dEnterria:2021ljz,Ren:2021prq, Cheung:2024qge,Bruggisser:2023npd} 
and other future colliders~\cite{Bauer:2018uxu,Bao:2022onq,Zhang:2021sio,Liu:2021lan,Han:2022mzp,Lu:2022zbe,Lu:2023ryd,Cheung:2024qge},
we find that most of the previous studies of ALP production channels at the LHC only focus on 
a single ALP effective operator in each process. However, people often overlook the potential for 
interference effects among different ALP operators except for the global analysis~\cite{Bruggisser:2023npd}. 
An interesting example in the SM is the associated production of the Higgs boson with a single top quark
\cite{Barger:2009ky,Biswas:2012bd,Biswas:2013xva,Farina:2012xp,Agrawal:2012ga,Ellis:2013yxa,Englert:2014pja,Chang:2014rfa}, in which the $HWW$ coupling interferes non-trivially with the top-Yukawa coupling and 
experimentally it can be disentangled. 
With the same spirit, we propose an alternative approach that takes into account two or more relevant ALP effective operators simultaneously in a single production process and explore
their interference effects. In particular, 
we study the ALP-$W^+ W^-$ and the ALP-$t\overline{t}$ couplings in the process $pp \to t j a$ at the LHC.

The process $pp\to tja$ as well as $pp\to j\gamma a$ and $pp\to t\bar ta$ have been considered in Ref.~\cite{Ebadi:2019gij}. However, interference effects among different ALP effective operators for these processes are not explored in that work. The process $pp\to j\gamma a$ involves the ALP-quark pair, ALP-gluon pair, ALP-$ZZ$, and ALP-$Z\gamma$ couplings. Similarly, the process $pp\to t\bar ta$ involves both ALP-$t\bar t$ and ALP-gluon pair couplings. Moreover, Ref.~\cite{Ebadi:2019gij} focuses only on $M_a < 100$ MeV, where the ALPs become invisible particles at the LHC. Therefore, further study on interference effects and heavy ALPs with prompt decays to SM particles is essential and complementary to exploring the properties of ALPs at the LHC.

In this work, we focus on the process $p p\rightarrow t j a$ with $a\to\gamma\gamma$, in which only the
ALP-$W^+ W^-$ and ALP-$t\overline{t}$ couplings are involved. Especially, the final state with two isolated photons and a top quark decaying semi-leptonically is considered for $25$ GeV $< M_a < 100$ GeV. Although this process is similar to the associated production of the Higgs boson with a single top quark, the relevant ALP coupling types are different from the Higgs boson ones. Therefore, the process $p p\rightarrow t j a$ can generate quite distinct predictions.
Our proposed approach allows us to explore novel ALP production processes that involve multiple 
ALP operators and investigate their interference effects.

The organization of this paper is as follows. In the next section, we describe the ALP interactions relevant to 
this study in the EFT framework. In Sec.~\ref{sec.3}, we explore the interference effects between the ALP-$W^+ W^-$ and ALP-$t\overline{t}$ couplings in $pp\to taX$ processes where "$X$" is the possible SM particles. In Sec.~\ref{sec.4}, we describe the experimental setup for discriminating the signal from the related SM backgrounds. We give
the numerical results and sensitivity reach of the ALP couplings in Sec.~\ref{sec.5}. Finally, we conclude in Sec.~\ref{sec.6}. 
\section{Theoretical setup}\label{sec.2}

In this study, the relevant ALP operators include the ALP-gauge boson pair and the ALP-top quark pair couplings, which start at
dimension-five. The $CP$-odd couplings of the ALP to the electroweak gauge boson fields are given by    
\begin{equation}\label{Eq.ALPLag}
\mathcal{L}_{\text{EW}}\supset-\frac{a}{f_a}\left( C_{WW} W_{\mu \nu}^i \tilde{W}^{i \mu \nu}+C_{BB} B_{\mu \nu} \tilde{B}^{\mu \nu}\right),
\end{equation}
where $i=1,2,3$ represents the $SU(2)$ index, and $\tilde{W}^{i \mu \nu}$ and 
$\tilde{B}^{\mu \nu}$ are the dual field strength tensors. 
Here the ALP field and its decay constant are represented by $a$ and $f_a$, respectively.
After transforming $W^i$ and $B$ to the physical fields $\gamma, Z, W^\pm$, the interactions in Eq.~(\ref{Eq.ALPLag}) can be written as 
\begin{equation}\label{Eq.ALPLag2}
\mathcal{L}_{\text{EW}}\supset  -\frac{1}{4}a\left( g_{a \gamma \gamma} F_{\mu \nu} \tilde{F}^{\mu \nu}+g_{a \gamma Z} F_{\mu \nu} \tilde{Z}^{\mu \nu}+g_{a Z Z} Z_{\mu \nu} \tilde{Z}^{\mu \nu}
   +g_{a W W} W_{\mu \nu} \tilde{W}^{\mu \nu}\right) \,,
\end{equation}
where $F_{\mu\nu}$, $W_{\mu\nu}$, and $Z_{\mu\nu}$ are the field strength tensors of 
the photon, $W^\pm$, and $Z$ bosons, respectively.
Thus, the dimensionful couplings of the photon and the electroweak gauge bosons to the ALP 
can be written in terms of $C_{WW}$ and $C_{BB}$~\cite{Brivio:2017ije,Georgi:1986df,Ren:2021prq,Cheung:2023nzg,Lu:2023ryd}, 
\begin{equation}\label{Eq.4}
g_{a\gamma\gamma}=\frac{4}{f_a}(C_{BB}c_w^2+C_{WW}s_w^2),
\end{equation}
\begin{equation}\label{Eq.5}
g_{aWW}=\frac{4}{f_a}C_{WW},
\end{equation}
\begin{equation}\label{Eq.6}
g_{aZZ}=\frac{4}{f_a}(C_{BB}s_w^2+C_{WW}c_w^2),
\end{equation}
\begin{equation}
g_{aZ\gamma }=\frac{8}{f_a}s_wc_w(C_{WW}-C_{BB}) \,,
\end{equation} 
where $c_w$ and $s_w$ are cosine and sine of the Weinberg angle 
that is related to the rotation between the electroweak fields and the physical fields as in
\begin{align*}
    W_{\mu}^{3}= c_w Z_\mu + s_w A_\mu, \quad B_{\mu}= -s_w Z_\mu + c_w A_\mu \,.
 \end{align*}

On the other hand, the ALP-top quark pair interaction is given by \cite{Esser:2023fdo}
\begin{equation}\label{Eq.7}
    \mathcal{L}_{at\overline{t}}= C_{a\phi}\frac{\partial_{\mu}{a}}{2f_a} (\bar{t}\gamma^{\mu}\gamma^5 t).
 \end{equation}
After applying the equation of motion, the above Lagrangian, Eq.~(\ref{Eq.7}), can be written as
\begin{equation}\label{Eq.ALPLag4}
    \mathcal{L}_{at\overline{t}}= -iC_{a\phi}\frac{m_{t}a}{f_a} (\bar{t}\gamma^5 t).
 \end{equation}
where $m_t$ is the top quark mass. As we can see, the ALP-quark pair coupling is proportional to the mass of the quark. Therefore, for the similar size of $C_{a\phi}/f_a$, the $at\overline{t}$ coupling can provide stronger interaction than other $aq\overline{q}$ couplings.
Equipped ourselves with these theoretical setups, we are now ready to discuss the interference effects between the $aW^+ W^-$ and the $at\overline{t}$ couplings in the process $pp \to tj a$.
\section{Production and Interference effects in $pp \to t a X$ processes}
\label{sec.3}

\begin{figure}[h!]
\centering
\includegraphics[width=5.5in]{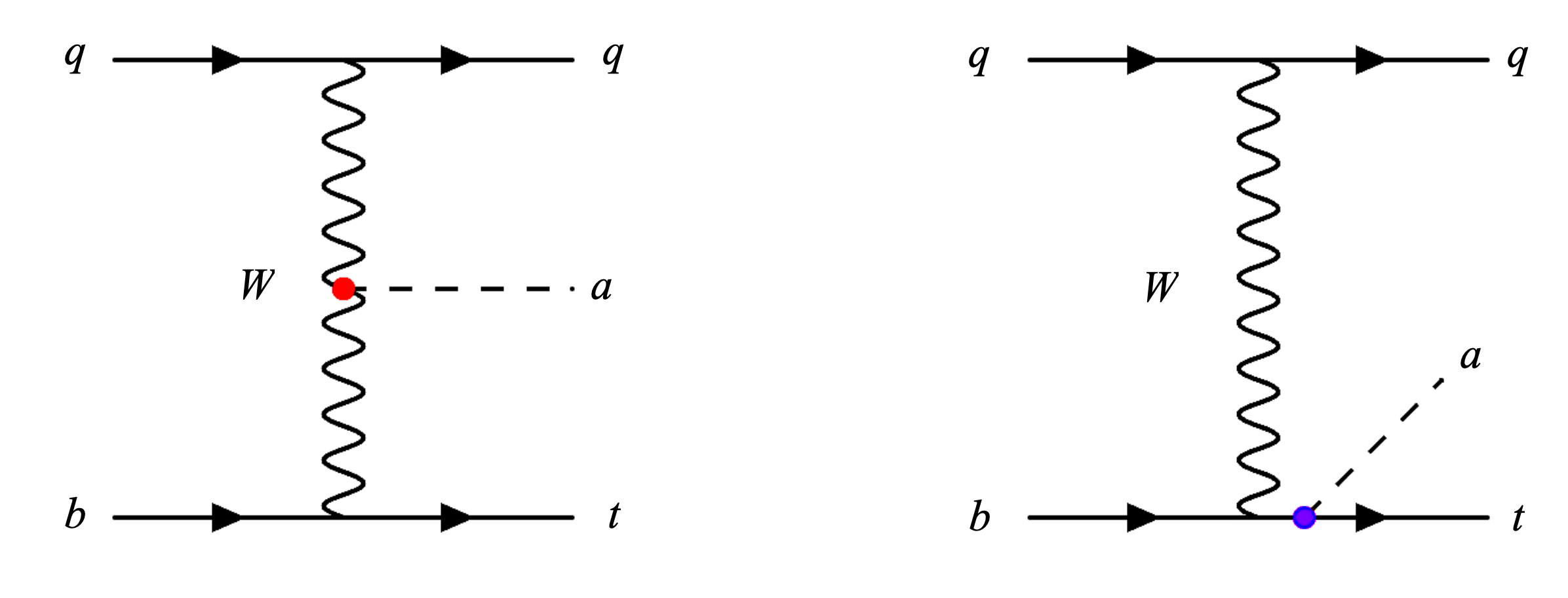}
\caption{\small \label{fig:feynman1} 
Two key contributing Feynman diagrams for the process $pp\to tj a$ at the LHC.
}
\end{figure}

\begin{figure}[th!]
\centering 
\includegraphics[width=0.8\textwidth]{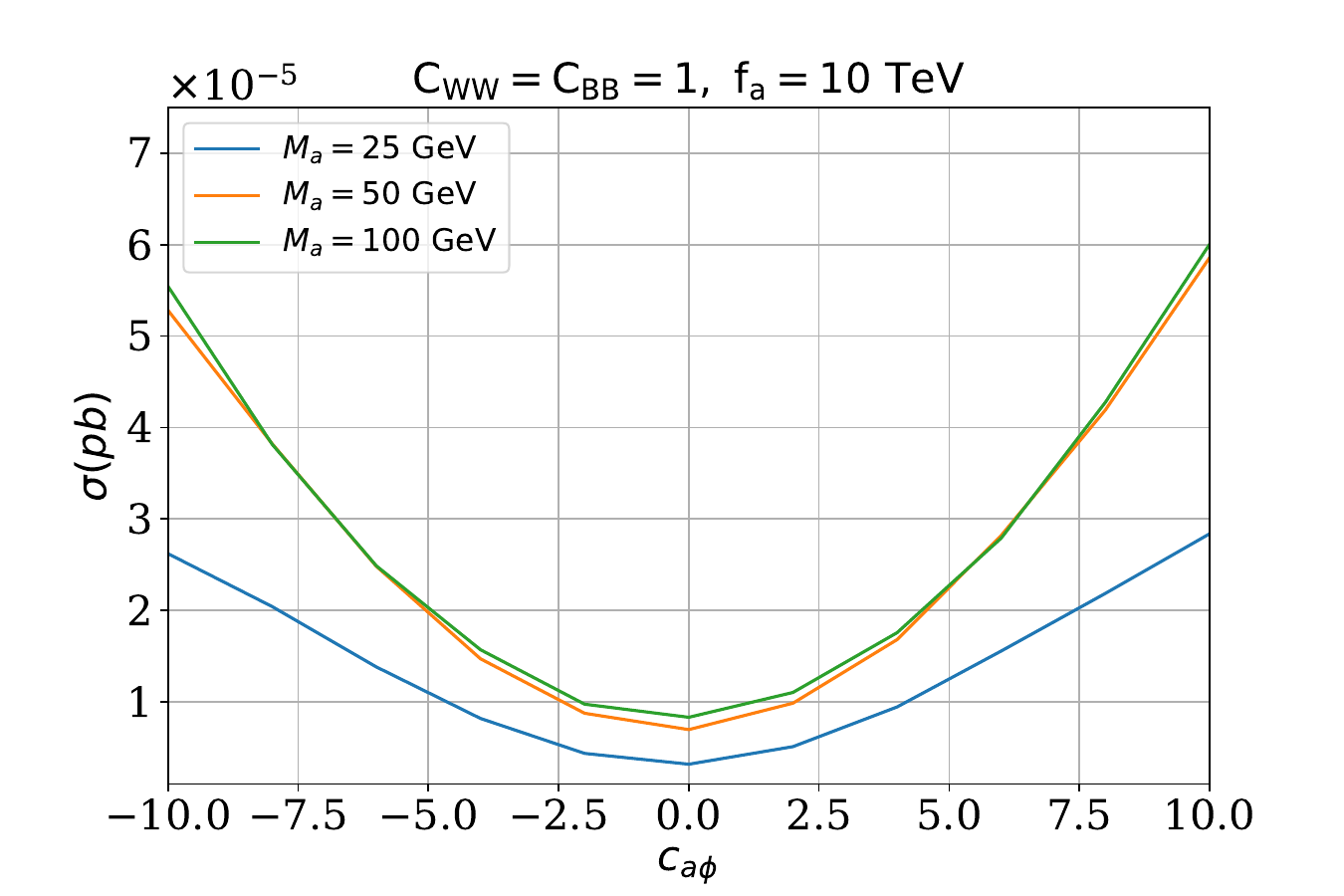} 
\caption{Production cross sections for the signal process $pp \to j~t~a$ with $a \to \gamma\gamma$ and $t \to bW,~W \to l\nu_l$ at the LHC ($\sqrt{s}=14~\rm TeV$) 
for $M_a=25, 50, 100$ GeV. We fix the $aW^+ W^-$ coupling by setting $C_{WW}=C_{BB}=1$ and
$f_a = 10$~TeV. The $at\overline{t}$ coupling $C_{a\phi}$ varies from $-10$ to $+10$.
}
\label{fig1}
\end{figure} 

\begin{figure}[h!]
\centering
\includegraphics[width=5.5in]{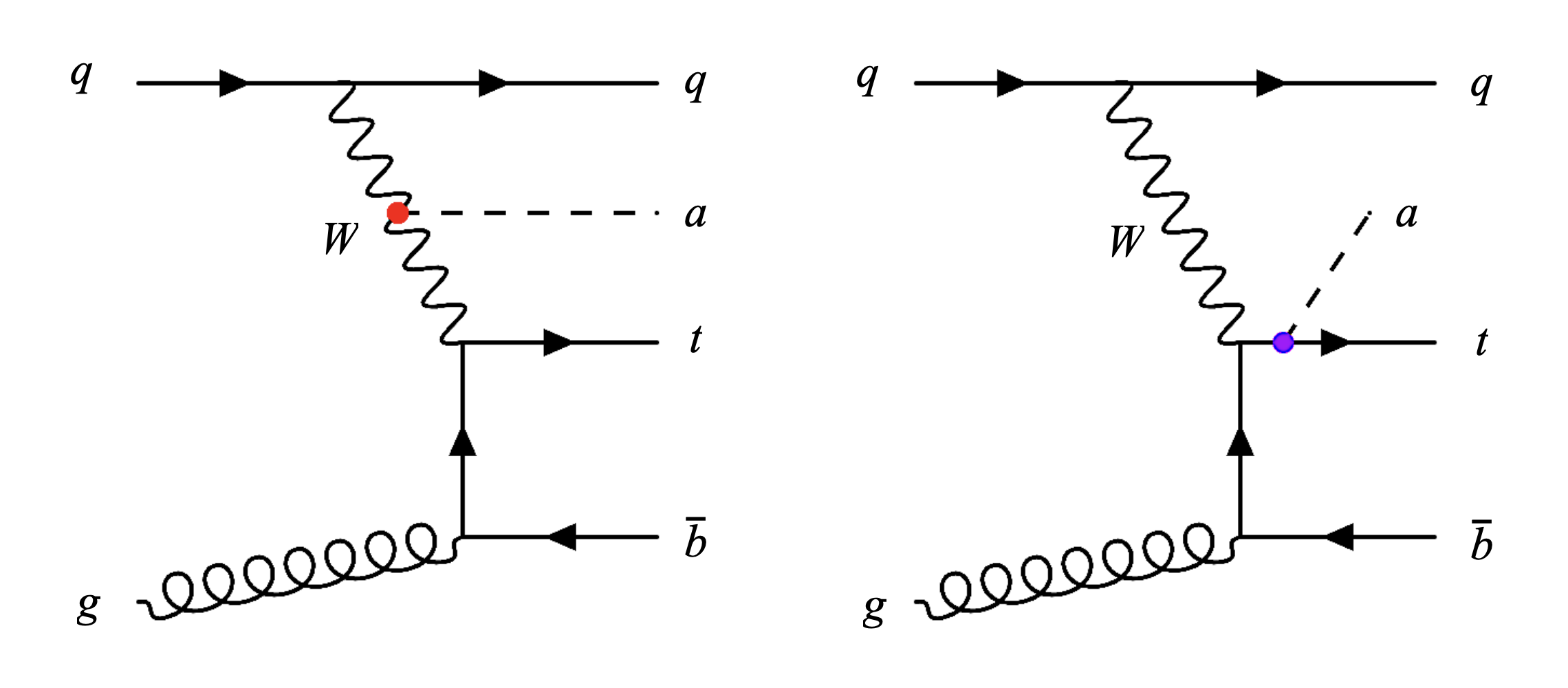}
\caption{\small \label{fig:feynman2} 
Two key contributing Feynman diagrams for the process $pp\to tjb a$ at the LHC.
}
\end{figure}

\begin{figure}[h!]
\centering
\includegraphics[width=5.5in]{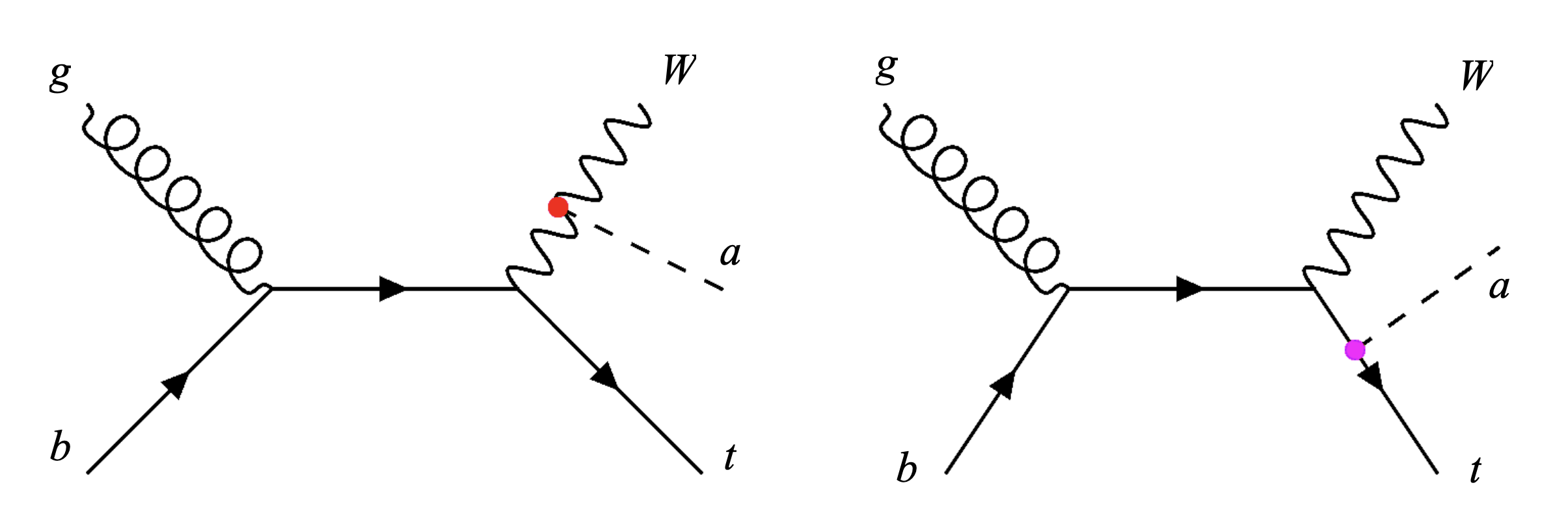}
\caption{\small \label{fig:feynman3} 
Two key contributing Feynman diagrams for the process $pp\to tW a$ at the LHC.
}
\end{figure}

\begin{figure}[h!]
\centering
\includegraphics[width=5.5in]{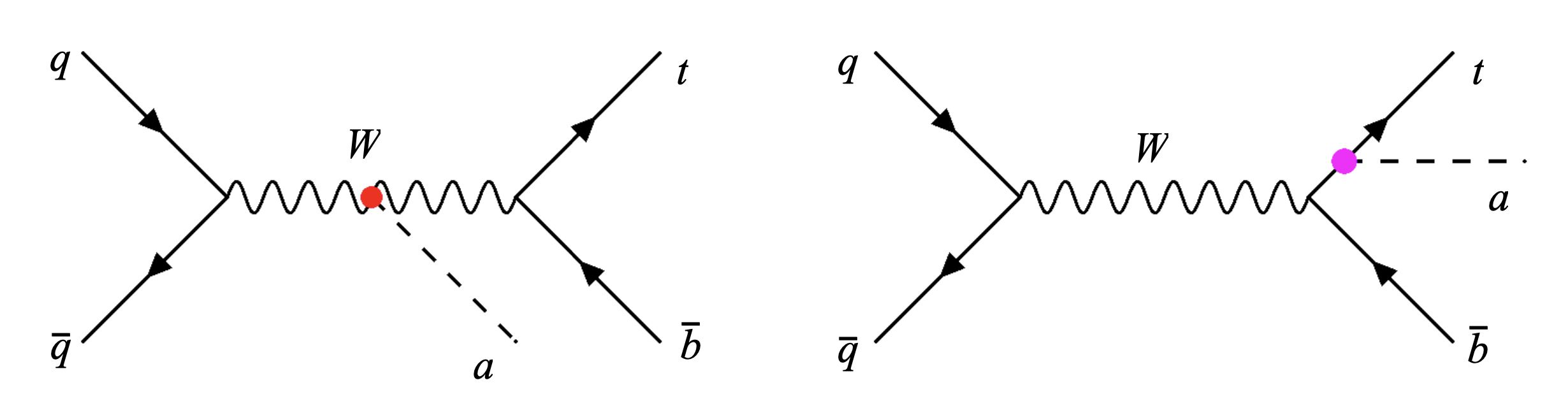}
\caption{\small \label{fig:feynman4} 
Two key contributing Feynman diagrams for the process $pp\to tb a$ at the LHC.
}
\end{figure}

In this section, our focus is on investigating the interference effects between 
the $aW^+ W^-$ and the $at\overline{t}$ operators in the process $pp\to taX$ where ``$X$" is the possible SM particles at the LHC
with $\sqrt{s}=14$ TeV.
Firstly, two key contributing Feynman diagrams for $pp \to t ja$ are shown in Fig.~\ref{fig:feynman1}. 
The ALP can bremsstrahlung off a $W$ propagator and also off a top-quark leg
\footnote{Here we do not involve the Feynman diagrams in which the ALP attached to the $b$ quark or 
light quarks since their contributions are much smaller than that from the 
$at\overline{t}$ coupling with the same $C_{a\phi}/f_a$ value.}. 
Thus, these two sets of diagrams can interfere. 

We apply \texttt{MadGraph5\_aMC@NLO}~\cite{MadGraph:2011} with the $5$-flavor scheme ($u,d,s,c,b$) to calculate the production cross sections for the process $pp \to t ja$ fixing the parameters: 
$C_{WW}=C_{BB}=1$ and $f_a = 10$~TeV, and vary the $at\overline{t}$ coupling labeled by $C_{a\phi}$ from $-10$ to $10$. 
We show in Fig.~\ref{fig1} the production cross sections for $M_a= 25, 50, 100$ GeV at the LHC ($\sqrt{s}=14~\rm TeV$),
including the branching ratios for $a \to \gamma\gamma$ and $t \to bW,~W \to l\nu_l$ ($l = e, \mu$). 
 
It is not difficult to see the interference effects when we look at the cross-section curves
at both ends ($-10$ and $+10$), although the effects are moderate at only about 10\% difference. 
Moreover, we have observed that the contribution from the $at\overline{t}$ interaction with $C_{a\phi}\sim 1$ is smaller than that from the $aW^+ W^-$ interaction with $C_{WW}=C_{BB}=1$ in this process. Meanwhile, the constraints for the $aW^+ W^-$ coupling are much stronger than that for the $at\overline{t}$ coupling as shown in Refs.~\cite{Alonso-Alvarez:2018irt,dEnterria:2021ljz,Esser:2023fdo}.

\begin{figure}[th!]
\centering 
\includegraphics[width=0.8\textwidth]{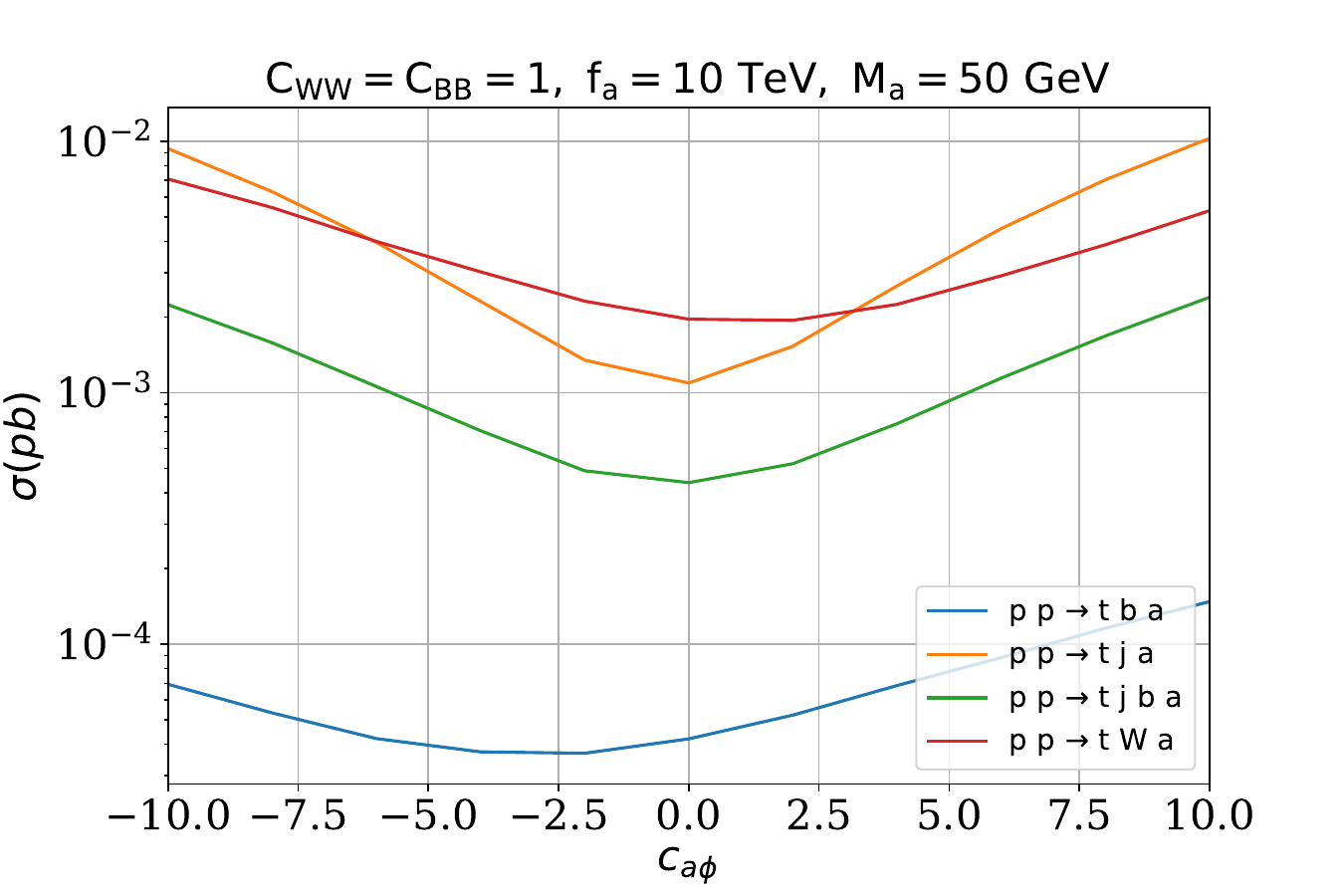} 
\caption{Production cross sections for the signal processes $pp \to t~j~a$, $pp\to t~j~b~a$, $pp\to t~W~a$, and $pp \to t~b~a$ at the LHC ($\sqrt{s}=14~\mathrm{TeV}$) for $M_a=50~\mathrm{GeV}$. We fix the ALP-gauge boson pair coupling by setting $C_{WW}=C_{BB}=1$ and $f_a = 10~\mathrm{TeV}$. The $at\overline{t}$ coupling $C_{a\phi}$ varies from $-10$ to $+10$.}
\label{fig1_xsection}
\end{figure}

Furthermore, we would like to discuss some other associated ALP production with a single top quark processes. The first one is $pp\to t~j~b~a$ (Fig.~\ref{fig:feynman2}), which can be regarded as a higher-order correction from $pp\to t~j~a$ when the $b$-quark is not tagged in the final state. To identify this process from $pp\to t~j~a$ and avoid the collinear divergence, the following cuts are applied to the $b$ and $j$ in the final state 
(note that the $P_{T_j}$ and $|\eta_j|$ cuts used in the 
signal-background analysis in the next section are different): 
\begin{equation}\label{Eq.tjba}
    P_{T_b} > 25~\text{GeV}, \quad |\eta_b| < 2.5, \quad P_{T_j} > 10~\text{GeV}, \quad |\eta_j| <5.
\end{equation}
The second and the third ones are $pp\to t~W~a$ (Fig.~\ref{fig:feynman3}) and $pp\to t~b~a$ (Fig.~\ref{fig:feynman4}) processes, respectively. In order to fairly compare the production cross-sections of these processes, we do not impose any cuts for them here, except for $pp\to t~j~b~a$ with the cuts in Eq.~(\ref{Eq.tjba}) to avoid the double-counting. The production cross-sections for the processes $pp \to t~j~a$, $pp\to t~j~b~a$, $pp\to t~W~a$, and $pp \to t~b~a$ at the LHC ($\sqrt{s}=14~\mathrm{TeV}$) for $M_a=50~\mathrm{GeV}$ are shown in Fig.~\ref{fig1_xsection}. We fix the ALP-gauge boson pair coupling by setting $C_{WW}=C_{BB}=1$ and $f_a = 10~\mathrm{TeV}$ and vary the $at\overline{t}$ coupling $C_{a\phi}$ from $-10$ to $+10$. 
Firstly, the shape of $pp\to t~j~b~a$ is similar to $pp\to t~j~a$, but the cross-section is smaller as we expect it 
to be a higher-order correction.
Secondly, $pp\to t~W~a$ is also a promising process which can show obvious interference effects. However, its cross-section is less sensitive to the variation of $C_{a\phi}$ than the one from $pp \to t~j~a$ and the decay modes of $W$ should be taken into account. Note that for the processes $pp\to tjba$ and $pp\to tWa$, the $agg$ coupling can also be included to study the interference effects among the $aW^+W^-$, $at\bar t$, and $agg$ couplings simultaneously. Finally, $pp \to t~b~a$ displays sizable interference effects as well, but its cross-section is much smaller than the other three processes. Therefore, we will stick with the process $pp \to t~j~a$ for the analysis in this study.
\section{Experimental setup and Simulations}
\label{sec.4}

In this section, we describe the calculation and experimental setup for 
discriminating the signal from dominant SM backgrounds. We show the event rates for the center-of-mass energy $\sqrt{s}=14$~TeV and integrated luminosities of $300~\rm fb^{-1}$ (current run) and 
$3000~\rm fb^{-1}$ (High-Luminosity LHC)\cite{Apollinari:2015bam}.

\subsection{Signal and relevant SM background processes}\label{sec.4.1}

The Monte Carlo simulations of signal and relevant SM background events are calculated utilizing 
\texttt{MadGraph5\_aMC@NLO}. The UFO model file of the ALP EFT framework 
(Eqs.~(\ref{Eq.ALPLag2}) and~(\ref{Eq.ALPLag4})) is employed for the signal event 
simulation~\cite{ALPmodel:2017}\footnote{This UFO model file is publicly accessible for download at 
\url{https://feynrules.irmp.ucl.ac.be/wiki/ALPsEFT}}. In our simulation, $10^4$ events are generated for the signal process and $10^5$ events for each 
SM background process. 

The subsequent steps involve parton showering and hadronization using \texttt{Pythia8} \cite{Sjostrand:2007gs},  and 
    detection simulations conducted with \texttt{Delphes3} \cite{deFavereau:2013fsa}, 
    incorporating the \texttt{ATLAS\_card.dat} for accuracy and consistency. Here a jet cone size $R = 0.4$ is employed for clustering jets using FastJet~\cite{Cacciari:2011ma} 
with the $anti$-$k_T$ algorithm~\cite{Cacciari:2008gp}.
The output root files from \texttt{Delphes3} are passed to the Python-based tool \texttt{uproot}~\cite{uproot5} for further analysis.

In order to investigate the final-state signature of two isolated photons, we focus on the 
ALP within a mass range spanning from $25~\rm GeV$ to $100~\rm GeV$\footnote{Two photons from the ALP decay will become too collimated to pass the isolation criteria when $M_a\lesssim 20$ GeV. In this situation, a photon-jet forms in the final state~\cite{Ren:2021prq,Lu:2022zbe,Cheung:2024qge} instead of two isolated photons.}. 
In this simulation, some specific benchmark values are assigned to the model parameters: 
$f_a=10~\rm TeV$, $C_{WW}=C_{BB}=1$, and a scan of $C_{a\phi}$ from $-10$ to $10$. 
A nonzero $C_{a\phi}$ together with $C_{WW}=1$ initiates ALP production from top bremsstralung, 
alongside with ALP production from $W$-boson fusion. 
In contrast, the choice of $C_{a\phi}=0$ prohibits the top bremsstralung into ALP. 
The variation $C_{a\phi}$ offers additional insights into the interference effects among these ALP operators.
We have already shown the interference effects in Fig.~\ref{fig1} in the last section.

The signal final state consists of the decay of the ALP and the top quark, as well as a 
hadronic jet. The dominant decay mode of the ALP is into a pair of isolated photons for the ALP mass
range from $25-100$~GeV for the setting $C_{WW}=C_{BB}=1$. We choose the semi-leptonic decay of the top quark:
$t\to W b,~\; W\to l \nu_l$ in this study. For such a final state, we consider two main SM backgrounds: 
(i) $p~p\to t~j~ \gamma~\gamma$ (labeled as BG1) and (ii) $p~p\to W~j~j~ \gamma~\gamma$ (labeled as BG2). 
Note that BG1 emerges as the predominant background in comparison to BG2. 
In order to estimate the sensitivity reach of the ALP coupling, we evaluate the total number of 
signal and background events at the LHC with a center-of-mass energy of $\sqrt{s}=14~\rm TeV$.
The total number of events is defined as:
\begin{equation}\label{eq:Event_rate}
        N_s, N_b = \sigma_{s,b} \times \frac{N_{\text{selected}}}{N_{\text{sim}}} \times \mathcal{L}\times \eta_{b-\text{tag}}\;,
\end{equation}
where $\sigma_b$ and $\sigma_s$ denote the cross-sections of background and signal events, respectively. 
The ratio $\frac{N_{\text{selected}}}{N_{\text{sim}}}$ represents the selection rate, and $\mathcal{L}$ 
is the integrated luminosity. 
The factor $\eta_{b-\text{tag} }$ represents the $b$-quark tagging efficiency
or $b$-mistag probability according to $b$ or $j$ in the context.

One may concern about the $t\bar t$-related backgrounds such as $t\bar t$, $t\bar t j$, or even
$t\bar t \gamma j$ when $j$'s are mis-tagged as photons. Since we have
applied isolation cuts among the photons, the jet, the $b$-jet, and lepton shown in 
Eq.~(\ref{B_cuts}), the mis-tag probability for $P_{j \to \gamma} \simeq 5 \times 10^{-4}$ 
\cite{ATLAS:2017muo}. Therefore, with such a small factor we do not expect 
these $t\bar t$ \-related backgrounds can affect significantly the sensitivity estimates.

\subsection{Event Selections }\label{sec.4.2}

In an effort to reduce these two main SM background events, we scrutinize the kinematic characteristics 
between the signal and background events, aiming to determine a suitable threshold. 
As discussed in the preceding section, we explore ALP masses ranging from $M_a=25~\rm GeV$ to 
$M_a=100~\rm GeV$. To illustrate interference effects, we select three benchmark values for the ALP mass, 
namely $M_a=10, 25,~\text{and}~100~\rm GeV$, and keeping $C_{W}=C_{B}=1$ fixed. 
Additionally, we examine the cases with $C_{a\phi}=-10,-5,0,5,10$. By observing variations 
in the number of events (as already depicted in Fig.~\ref{fig1}, where the event rate 
quantifies the interference effect) for couplings with the same magnitude but 
opposite signs, we can readily discern the interference effect.
Finally, the results for the integrated luminosities 300 fb$^{-1}$ 
(current run) and  3000 fb$^{-1}$  (High-Luminosity LHC) will be shown in the next section. 

\begin{figure}[th!]
\centering
\includegraphics[width=6in]{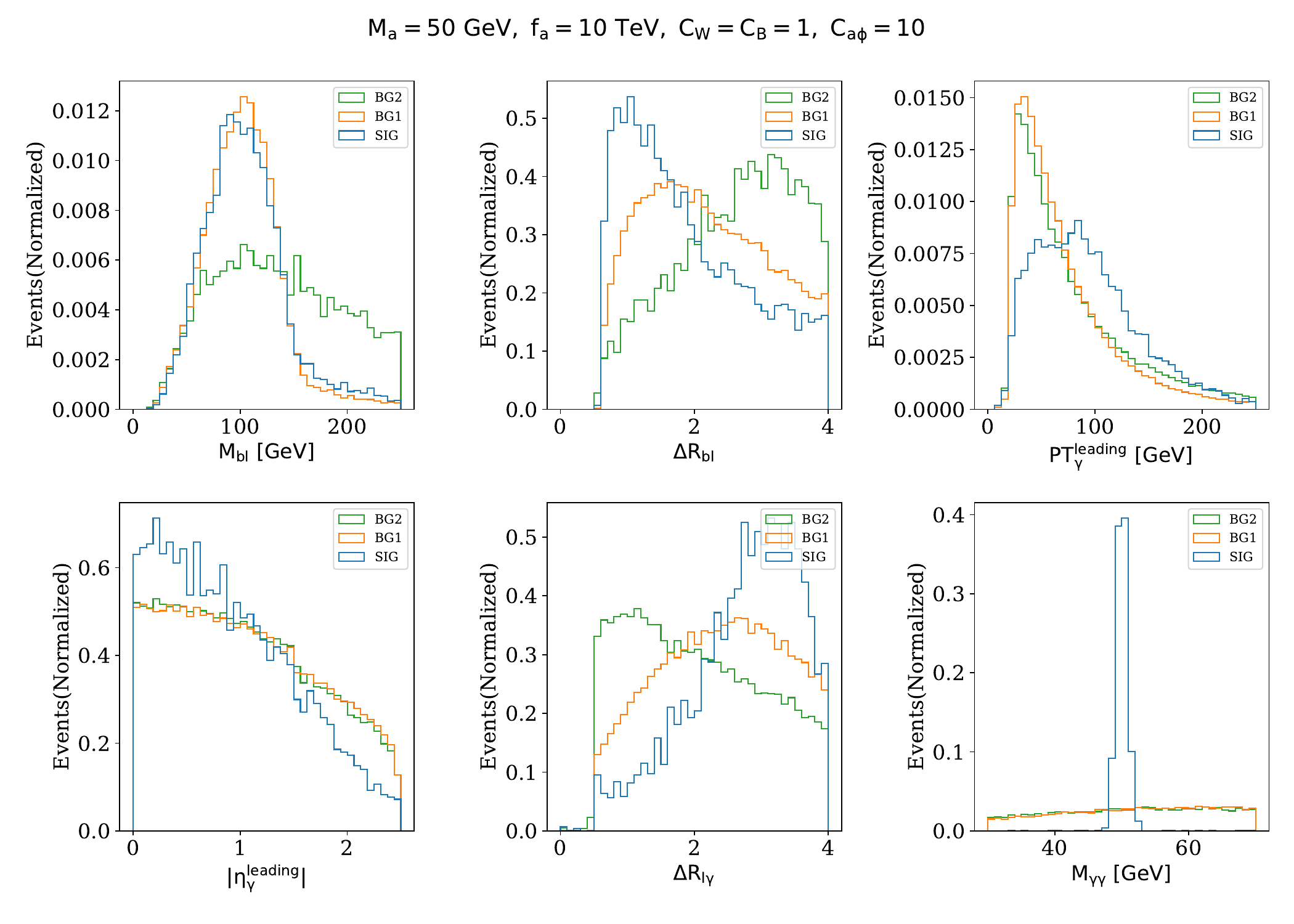}
\caption{Kinematical distributions for the signal with $M_a=50$ GeV and 
two main SM backgrounds BG1 and BG2. Here $f_a= 10$ TeV and $C_{WW}=C_{BB}=1$ are fixed.
The ``leading" in $P^{\rm leading}_{T_\gamma}$ and $|\eta^{\rm leading}_\gamma|$ refers to the photon with the highest transverse momentum. 
\label{fig:dist}
}
\end{figure}

Various kinematical distributions for the signal case of $M_a=50$ GeV along with two main SM backgrounds at detector level are shown in Fig.~\ref{fig:dist}. The signal cases for $M_a=25$ GeV and $100$ GeV show similar behavior. As an initial requirement for event selection, we applied a set of cuts on the transverse momentum \(P_T\) and rapidity \(|\eta|\) of the final-state particles. We refer to these cuts as ``Basic Cuts''. The Basic Cuts are given below in Eq.~(\ref{B_cuts}):  
\begin{equation}\label{B_cuts}
\begin{aligned}
& \textbf{Basic Cuts:} \\
& P_{T_b} > 25~\text{GeV}, \quad |\eta_b| < 2.5, \quad P^{\rm lead}_{T_j} > 25~\text{GeV}, \quad |\eta^{\rm lead}_j| < 2.5, \\
& P_{T_l} > 25~\text{GeV}, \quad |\eta_l| < 2.5, \quad P^{\rm lead}_{T_\gamma}~\&~ P^{\rm sub-lead}_{T_\gamma} > 20~\text{GeV}, \quad |\eta^{\rm lead}_\gamma| ~\&~ |\eta^{\rm sub-lead}_\gamma| < 2.5, \\
& \Delta R_{ij} > 0.4~\text{for} ~i,j = \gamma, l, j, b~ (\text{except}~\Delta R_{\gamma\gamma} > 0.3). 
\end{aligned}
\end{equation}
Here ``lead" and ``sub-lead" refer to the leading and sub-leading orders according
to $P_T$. Additionally, the $b$-quark tagging efficiency and mistag probability are adopted from the ATLAS template in \texttt{Delphes3}.

We impose the following cuts for event selection: \( M_{bl} < 150 \) GeV, ensuring the invariant mass of the system consisting of the \( b \)-jet and the charged lepton from the top-quark decay is less than 150 GeV. This helps isolate events where the \( b \)-jet and charged lepton originate from the top-quark decay. The cut \( \Delta R_{bl} < 2.5 \) ensures that the angular separation between the \( b \)-jet and the charged lepton (in \( \eta \)-\( \phi \) space) is less than 2.5, ensuring the \( b \)-jet and charged lepton are sufficiently close, as expected from top-quark decays. We require \( P^{\rm leading}_{T_\gamma} > 60 \) GeV, ensuring the transverse momentum of the leading (highest \( p_T \)) photon is greater than 60 GeV, to select high-energy photons 
from the ALP decay.
The cut \( |\eta^{\rm leading}_{\gamma}| < 1.7 \) ensures the pseudorapidity of the leading photon is within 1.7, selecting photons within the central detector region where detection efficiency is higher.  The cut on \( \Delta R_{l\gamma} > 2.0 \) ensures the angular separation between the charged lepton and the photon (in \( \eta \)-\( \phi \) space) is greater than 2.0, ensuring the charged lepton and photon are well separated, reducing the background from misidentified charged leptons and photons. Finally, we impose the invariant-mass window cut on the
diphoton from the ALP decay: $|M_{\gamma\gamma}-M_a| < 5$ GeV. As shown in Fig.~\ref{fig:dist}, the ALP mass window within 10 GeV is broad enough to encompass the major part of this resonance in the invariant diphoton mass distribution. Our cut on $|M_{\gamma\gamma}-M_a| < 5$ GeV could be further refined in a real experimental bump hunt analysis. Since the chosen ALP mass window reflects the energy measurement precision of the relevant detectors, such an analysis is beyond the scope of this study.

We summarize the above cuts for the signal and background event selections:
\begin{itemize} 
    \item \textbf{Basic Cuts} in Eq.~(\ref{B_cuts}), 
    \item $M_{bl}<150~\rm GeV$, 
    \item $\Delta R_{bl}<2.5$, 
    \item $P^{\rm leading}_{T_\gamma}>60~\rm GeV$,  
    \item $|\eta^{\rm leading}_{\gamma}|<1.7$, 
    \item $\Delta R_{l\gamma}>2.0$, 
    \item $|M_{\gamma\gamma}-M_a| < 5$ GeV. 
\end{itemize}
The cut-flow tables for $M_a=25, 50$, and 100 GeV are given in 
Tables~\ref{tab:cutflow_25}, \ref{tab:cutflow_50} \ref{tab:cutflow_100}, 
respectively. We found that the ALP invariant-mass window cut is the strongest one to reduce events from both BG1 and BG2 but keep the signal events.

\begin{table}
\begin{ruledtabular}
   \begin{tabular}{l|ccccccc}
       \textbf{Cut} & \textbf{BG1} & \textbf{BG2} & \multicolumn{5}{c}{\textbf{Signal}} \\
        \cline{4-8}
        & & & \textbf{$C_{a\phi}=-10$} & \textbf{$C_{a\phi}=-5$} & \textbf{$C_{a\phi}=0$} & \textbf{$C_{a\phi}=5$} & \textbf{$C_{a\phi}=10$} \\
        \hline
        \textbf{Basic Cuts} & 404.35& 8.95& 4.25 & 1.64 & 0.35 & 1.85 & 4.67 \\
        $M_{bl}<150~\rm GeV$ & 371.85 & 3.23 & 3.88 & 1.50 & 0.32 & 1.69 & 4.24 \\
        $\Delta R_{bl}<2.5$ & 219.2 & 1.72 & 2.38 & 0.97 & 0.18 & 1.00 & 2.58 \\
        $P^{\rm leading}_{T_\gamma}>60~\rm GeV$ & 135.46 & 1.17 & 0.92 & 0.34 & 0.09 & 0.40 & 1.14 \\
        $|\eta^{\rm leading}_{\gamma}|<1.7$& 111.89 & 1.00 & 0.83 & 0.30 & 0.07 & 0.35 & 0.93 \\
        $\Delta R_{l\gamma}>2.0$ & 87.47 & 0.66 & 0.69 & 0.23 & 0.05 & 0.29 & 0.82 \\
        $20~\rm GeV < M_{\gamma\gamma} < 30~\rm GeV$ & 0.74 & 0.00 & 0.69 & 0.23 & 0.05 & 0.29 & 0.80 \\
    \end{tabular}
\end{ruledtabular}
    \caption{Cutflow table for the SM backgrounds 
    (BG1: $p~p\to t~j~ \gamma~\gamma$ and BG2: $p~p\to W~j~j~ \gamma~\gamma$),
    and the signal: $p~p\to j~t~a$ with various $C_{a\phi}$ couplings. Here we set 
    $M_{a}=25~\text{GeV},~f_a=10~\text{TeV},~C_{WW}=C_{BB}=1 $, and $C_{a\phi} = -10, -5, 0, 5, 10$ GeV.
    \textbf{Basic Cuts} in the first row denotes the total number of events passed the cuts coded in Eq.~(\ref{B_cuts}).
    The number of events are calculated by Eq.~(\ref{eq:Event_rate}) and integrated luminosity 
    is set to $\mathcal{L}=3000~\rm fb^{-1}$. }
    \label{tab:cutflow_25}
\end{table}

\begin{table}
\begin{ruledtabular}
   \begin{tabular}{l|ccccccc}
       \textbf{Cut} & \textbf{BG1} & \textbf{BG2} & \multicolumn{5}{c}{\textbf{Signal}} \\
        \cline{4-8}
        & & & \textbf{$C_{a\phi}=-10$} & \textbf{$C_{a\phi}=-5$} & \textbf{$C_{a\phi}=0$} & \textbf{$C_{a\phi}=5$} & \textbf{$C_{a\phi}=10$} \\
        \hline
        \textbf{Basic Cuts} & 404.35 & 8.95 & 18.52 & 6.46 & 2.07 & 7.36 & 20.84 \\
        $M_{bl}<150~\rm GeV$ & 371.85 & 3.23 & 16.09 & 5.72 & 1.88 & 6.59 & 18.26 \\
        $\Delta R_{bl}<2.5$ & 219.2 & 1.72 & 11.81 & 4.11 & 1.26 & 4.42 & 12.35 \\
        $P^{\rm leading}_{T_\gamma}>60~\rm GeV$ & 135.46 & 1.17 & 9.86 & 3.54 & 1.17 & 3.81 & 10.51 \\
        $|\eta^{\rm leading}_{\gamma}|<1.7$ & 111.89 & 1.0 & 8.84 & 3.18 & 0.98 & 3.36 & 9.4 \\
        $\Delta R_{l\gamma}>2.0$ & 87.47 & 0.66 & 7.99 & 2.72 & 0.83 & 3.05 & 8.69 \\
        $45~\rm GeV < M_{\gamma\gamma} < 55~\rm GeV$ & 3.2 & 0.02 & 7.91 & 2.68 & 0.82 & 3.01 & 8.62 \\
    \end{tabular}
\end{ruledtabular}
    \caption{Cutflow table for the SM backgrounds 
    (BG1: $p~p\to t~j~ \gamma~\gamma$ and BG2: $p~p\to W~j~j~ \gamma~\gamma$),
    and the signal: $p~p\to j~t~a$ with various $C_{a\phi}$ couplings. Here we set 
    $M_{a}=50~\text{GeV},~f_a=10~\text{TeV},~C_{WW}=C_{BB}=1 $, and $C_{a\phi} = -10, -5, 0, 5, 10$ GeV.
    \textbf{Basic Cuts} in the first row denotes the total number of events passed the cuts coded in Eq.~(\ref{B_cuts}).
    The number of events are calculated by Eq.~(\ref{eq:Event_rate}) and integrated luminosity 
    is set to $\mathcal{L}=3000~\rm fb^{-1}$. }
    \label{tab:cutflow_50}
\end{table}

\begin{table}
\begin{ruledtabular}
   \begin{tabular}{l|ccccccc}
       \textbf{Cut} & \textbf{BG1} & \textbf{BG2} & \multicolumn{5}{c}{\textbf{Signal}} \\
        \cline{4-8}
        & & & \textbf{$C_{a\phi}=-10$} & \textbf{$C_{a\phi}=-5$} & \textbf{$C_{a\phi}=0$} & \textbf{$C_{a\phi}=5$} & \textbf{$C_{a\phi}=10$} \\
        \hline
        \textbf{Basic Cuts} & 404.35 & 8.95 & 27.3 & 10.31 & 4.22 & 11.07 & 31.21 \\
        $M_{bl}<150~\rm GeV$ & 371.85 & 3.23 & 23.67 & 8.91 & 3.75 & 9.58 & 27.16 \\
        $\Delta R_{bl}<2.5$ & 219.2 & 1.72 & 17.92 & 6.75 & 2.73 & 7.04 & 20.63 \\
        $P^{\rm leading}_{T_\gamma}>60~\rm GeV$ & 135.46 & 1.17 & 17.29 & 6.49 & 2.7 & 6.83 & 19.84 \\
        $|\eta^{\rm leading}_{\gamma}|<1.7$ & 111.89 & 1.0 & 15.63 & 5.8 & 2.3 & 6.07 & 17.91 \\
        $\Delta R_{l\gamma}>2.0$ & 87.47 & 0.66 & 13.98 & 5.1 & 1.92 & 5.49 & 16.77 \\
        $95~\rm GeV < M_{\gamma\gamma} < 105~\rm GeV$ & 5.96 & 0.04 & 13.93 & 5.05 & 1.9 & 5.46 & 16.7 \\
    \end{tabular}
\end{ruledtabular}
    \caption{Cutflow table for the SM backgrounds 
    (BG1: $p~p\to t~j~ \gamma~\gamma$ and BG2: $p~p\to W~j~j~ \gamma~\gamma$),
    and the signal: $p~p\to j~t~a$ with various $C_{a\phi}$ couplings. Here we set 
    $M_{a}=100~\text{GeV},~f_a=10~\text{TeV},~C_{WW}=C_{BB}=1 $, and $C_{a\phi} = -10, -5, 0, 5, 10$ GeV.
    \textbf{Basic Cuts} in the first row denotes the total number of events passed the cuts 
    coded in Eq.~(\ref{B_cuts}).
    The number of events are calculated by Eq.~(\ref{eq:Event_rate}) and integrated luminosity 
    is set to $\mathcal{L}=3000~\rm fb^{-1}$.}
    \label{tab:cutflow_100}
\end{table}

\section{Numerical results}\label{sec.5}

\begin{figure}[h]
\centering 
\includegraphics[width=\textwidth]{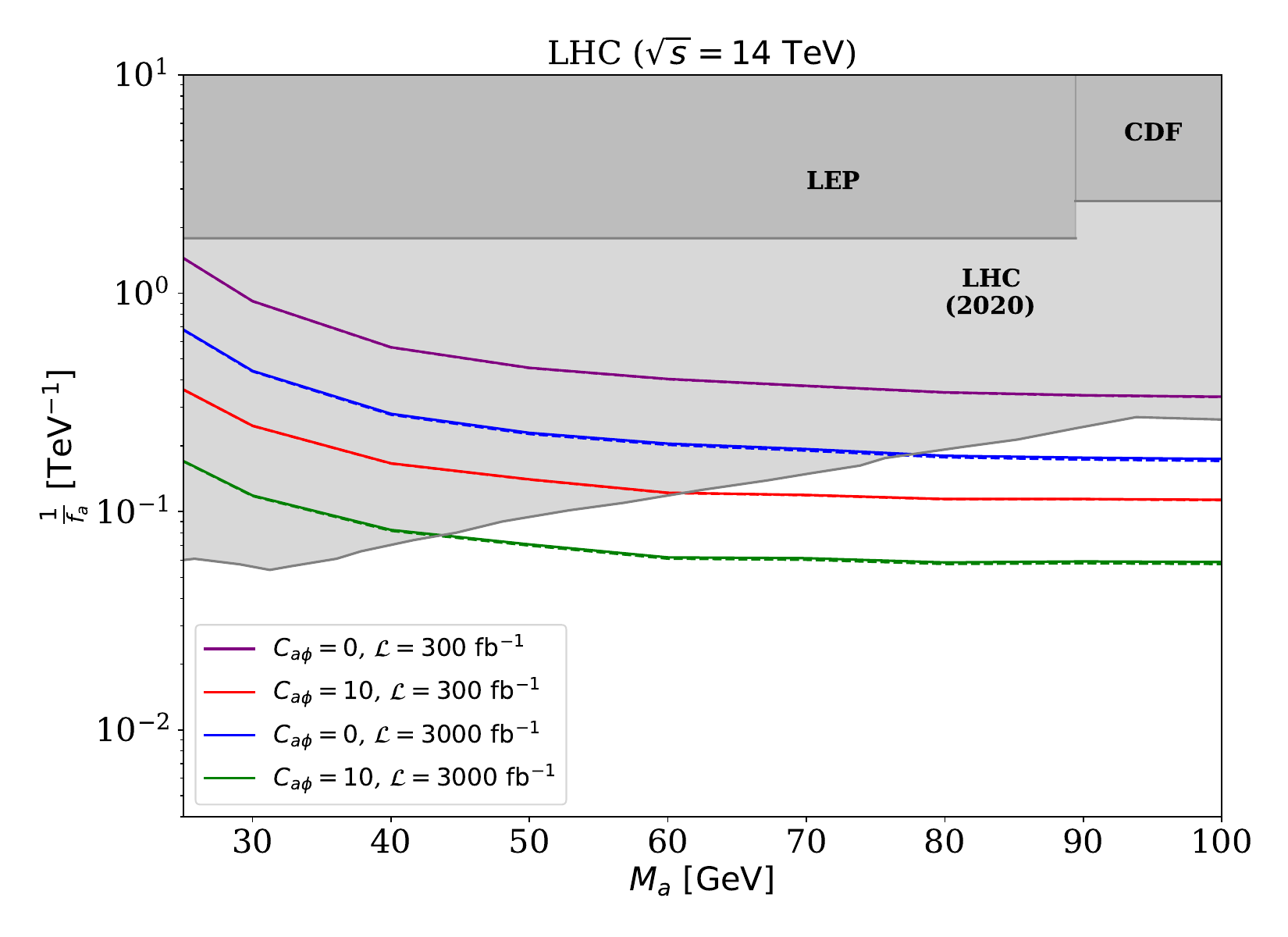}
\caption{\label{fig:results}Exclusion regions at 95\% confidence level (C.L.) for the ALP cutoff scale 
$f_a$ derived from the process $pp \to j~t~a$ followed by $a \to \gamma\gamma$ and 
$t \to b l\nu_l$ at the LHC with $\sqrt{s}=14$ TeV. This analysis is conducted with different choices of 
$C_{a\phi}=0,\,10$ and under two sets of integrated luminosities,
$\mathcal{L}=300~\rm fb^{-1}$ and $ 3000~\rm fb^{-1}$. 
Solid lines in the plot represent sensitivity curves with a $10\%$ systematic 
uncertainty, while the dashed lines depict the curves without incorporating 
systematic uncertainty. 
The gray areas represent the existing limits from LEP~\cite{L3:1994shn, OPAL:2002vhf}, CDF~\cite{CDF:2013lma}, 
and LHC~\cite{Clark:2011zza, Masciovecchio:2016gil}. 
}
\label{Sen_dis}
\end{figure} 

After imposing the event selections provided in the last section, the number
of signal events are comparable to, if not larger than, the background events.
It is thus meaningful to calculate the significance of the signal and set limits 
on the cutoff scale $f_a$. The relation between the ALP signal events, $N_s$, and the ALP cutoff scale, $f_a$, is given as follows:
\begin{equation}\label{Eq. s_f}
    N_s \propto f_a^{-2} \;.
\end{equation}
Therefore, we can rescale the factor $f_a$ to match the expected signal events $N_s$.
The significance of the signal is given by \cite{Cheung:2023nzg}
\begin{equation}\label{Eq:sign}
     Z = \sqrt{2\left[(N_s+N_b)\ln\left(\frac{(N_s+N_b)(N_b+\sigma_B^2)}{{N_b}^2+(N_s+N_b)\sigma_B^2}\right) - \frac{N_b^2}{\sigma_B^2}\ln\left(1+\frac{\sigma_B^2 N_s}{N_b(N_b+\sigma_B^2)}\right)\right]} \;,
\end{equation}
where $N_s$, $N_b$ are the number of signal and background events, and $\sigma_B$ is
the systematic uncertainty in background estimation, which is taken to be zero and
$0.1 N_b$ in the presentation.
The $95\%$ confidence level (C.L.) sensitivity curves for the ALP cutoff scale 
$f_a$ to the ALP mass $M_a$ are obtained by requiring the significance $Z > 2$.

In Fig.~\ref{Sen_dis}, we show $95\%$ C.L. exclusion region 
for the ALP cutoff scale, $f_a$. This exclusion region is obtained through the
process $pp \to j~t~a$ with $a \to \gamma\gamma$ and $t \to bW,~W \to l\nu_l$ at the LHC with $\sqrt{s}=14$ TeV. In the figure, we fix $C_{WW}=C_{BB}=1$ and set $C_{a\phi}=0$ and $C_{a\phi}=10$ as two benchmark points. Two different sets of integrated luminosities, $\mathcal{L}=300~\rm fb^{-1}$ and 
$3000~\rm fb^{-1}$ are plotted. 
In the plot, solid lines denote sensitivity curves accounting for a $10\%$ systematic 
uncertainty, while dashed lines represent curves without incorporating systematic 
uncertainty. We can find that the systematic uncertainty only slightly changes the predictions of future bounds in this study. It indicates that our results are robust against possible uncertainties.
The gray shaded areas correspond to the existing limits from LEP 
\cite{L3:1994shn, OPAL:2002vhf}, CDF \cite{CDF:2013lma}, and
LHC \cite{Clark:2011zza, Masciovecchio:2016gil}. 
The lines in Fig.~\ref{Sen_dis} unmistakably demonstrate that sensitivity 
curves with non-zero values of $C_{a\phi}$ can yield better limits on $f_a$
than those with $C_{a\phi}=0$. It can also be observed in Fig.~\ref{fig1} in which the minimal production cross section for the signal process appears at $C_{a\phi}=0$ and it enhances as the $|C_{a\phi}|$ increases.

Specifically, the sensitivity reach corresponding to $C_{a\phi}=10$ and 
$\mathcal{L}= 300\;(3000)~\rm fb^{-1}$ (depicted by the red (green) line) 
exhibits enhanced sensitivity compared to current constraints for 
$M_a >60$ GeV ($M_a >45$ GeV). 
Even in the absence of the $C_{a\phi}$ coupling ($C_{a\phi}=0$), the HL-LHC 
($\mathcal{L}=3000~\rm fb^{-1}$) can impose the sensitivity on $f_a$ (indicated by the blue curve) for $M_a>80~\rm GeV$.

\section{Conclusions}\label{sec.6}

In summary, we have presented a novel approach that considers simultaneous
presence of two or more ALP interaction operators in a single process at the LHC. 
In particular, we demonstrated the interference effects of the 
ALP-gauge boson pair and the ALP-top quark pair couplings in the process $pp\to t ja$ as shown in Fig.~\ref{fig1}. 
Through a detailed analysis of $p p\rightarrow t j a$,
followed by $a\rightarrow\gamma\gamma$ and semi-leptonic decay of the top quark,
as a case study, we demonstrated the efficacy of this approach in constraining ALP 
interactions as well as their interference within a single process.

Our findings indicate that the sensitivity of the ALP cutoff scale $f_a$ could 
potentially reach down to the values around $1/f_a \sim 5 \times 10^{-2}\;
{\rm TeV}^{-1}$ for the ALP masses ranging from $25$ GeV to $100$ GeV at the HL-LHC as shown in Fig.~\ref{fig:results}. It indicates that some uncovered parameter space can be further explored from the process $pp\rightarrow tja$ in the near future. Furthermore, once the absolute size of the ALP-gauge boson pair and the ALP-top quark pair couplings can 
be pinned down by other ALP production channels, 
this process can provide extra information about the relative sign (or phase) between two coefficients 
of ALP interaction operators. In our case, we observe a positive interference between the processes in Fig.~\ref{fig:feynman1} which is visible from the fact that the cross-section increases with $C_{a\phi}$.


\section*{Acknowledgment}

The work of K.Cheung, P.Sarmah, and C.J. Ouseph is supported by 
the Taiwan NSTC with grant no. MoST-110-2112-M-007-017-MY3. The work of C.-T. Lu is supported by the Special funds for postdoctoral overseas recruitment, Ministry of Education of China (No.~164080H0262403).

\bibliography{paper}

\end{document}